\documentclass{ws-procs975x65}

\newcommand{\be}{\begin{equation}}
\newcommand{\ee}{\end{equation}}
\newcommand{\ba}{\begin{eqnarray}}
\newcommand{\ea}{\end{eqnarray}}
\def\bs{\begin{subequations}}
\def\es{\end{subequations}}

\newcommand{\rt}{\tilde{r}}
\newcommand{\vpt}{\tilde{\varphi}}

\begin{document}

\title{\uppercase{Fixed functionals in Asymptotically Safe Gravity}
\footnote{Talk given by F.S. at the 13th Marcel Grossmann Meeting, Stockholm, 1-7 July 2012.}
}

\author{M.\ Demmel, F.\ Saueressig and O.\ Zanusso}

\address{PRISMA Cluster of Excellence \& Institute of Physics (THEP), \\
University of Mainz, D-55128 Mainz, Germany%
}
\vspace{-10pt}
\begin{abstract}
We summarize the status of constructing fixed functionals within the $f(R)$-truncation of Quantum Einstein Gravity in three spacetime dimensions. Focusing on curvatures much larger than the IR-cutoff scale, it is shown that the fixed point equation admits three different scaling regimes: for classical and quantum dominance the equation becomes linear and has power-law solutions, while the balanced case gives rise to a generalized homogeneous equation whose order is reduced by one and whose solutions are non-analytical.
\end{abstract}

\keywords{Quantum gravity; Asymptotic Safety; Fixed Point}

\bodymatter\bigskip
\section{Introduction}
During the last decade Weinberg's Asymptotic Safety conjecture \cite{Weinberg1979} has undergone an extensive transformation from a curiosity to a serious candidate theory for quantum gravity.
The heart of the construction is a non-Gaussian fixed point (NGFP) in the renormalization group (RG) flow,
which controls the behavior of the theory at high energies and renders it predictive and free from unphysical divergences \cite{Niedermaier2006}.
One of the central technical ingredients for testing Asymptotic Safety is the functional renormalization group equation for the gravitational effective average action $\Gamma_k$ \cite{Reuter1998b}.
By now, this equation has been used to demonstrate the existence of the NGFP in a wide variety of approximations and coarse-graining schemes \cite{Reuter:2012id}.
Since a finite-dimensional truncation of $\Gamma_k$ will always yield a finite number of relevant operators, testing the predictive power of Asymptotic Safety requires the study of truncations that involve infinitely many coupling constants.
This program is currently implemented by considering the gravitational RG-flow of $f(R)$-type truncations\cite{Codello:2007bd,Machado:2007ea} where the gravitational part of $\Gamma_k$ is of the form
\be\label{truncation}
\begin{split}
 \Gamma_k^{\rm grav}[g] &= \int {\rm d}^dx\,\sqrt{g}\,f_k\!\left(R\right)\,,
\end{split}
\ee
with $g_{\mu\nu}$ being the spacetime metric and $f_k$ an arbitrary function of the curvature scalar $R$. The partial differential equations (PDEs) resulting from substituting this ansatz into the full flow equation have recently been studied in the particular cases of three \cite{Demmel:2012ub} and four spacetime dimensions \cite{Benedetti:2012dx,Dietz:2012ic,Benedetti:2013jk}.
These PDEs encode the dependence of the function $f_k$ on the RG-scale $k$ and thus store the information of the flow of infinitely many couplings including the effective Newton's constant.
Fixed functionals, providing the generalization of the NGFP to the case of flowing functions, then correspond to $k$-independent solutions of these PDEs and the number of relevant deformations of the fixed points can be found by studying linear deformations around such a fixed functional.

\section{The flow of $f_k(R)$ in three dimensions}
In the context of $d=3$ conformally reduced gravity, where only the conformal mode of the graviton is dynamical, the flow of $f(R)$ was recently studied in [\refcite{Demmel:2012ub}].
The non-linear PDE governing the scale-dependence of the dimensionless function $ \varphi_k(r) \equiv k^{-3} f_k(k^2r)$, $\dot{\varphi}_k\equiv k\partial_k\varphi_k$, reads
\be\label{pdgl}
\dot{\varphi}_k
+ 3\varphi_k - 2r\varphi'_k
=
\frac{1}{\pi^2}\left( 1 + \frac{r}{6}  \right)^\frac{3}{2}
\frac{c_1\varphi'_k +  c_2 \varphi''_k +  c_3\dot{\varphi}'_k + c_4\left(\dot{\varphi}''_k-2r\varphi'''_k\right)}{3\varphi_k+4(1-r)\varphi'_k + 4\left(2-r\right)^2\varphi''_k}\,.
\ee
The coefficients are polynomials in the dimensionless curvature $r$
\be\label{ccoeff}
\begin{split}
\begin{aligned}
c_1 &= \frac{1}{45} \left( 36 + r \right) \, , &  c_2 =& \frac{2}{189} \left(5 r^2 - 234 r + 432 \right) \, , \\
c_3 &= \frac{1}{45} \left( 6 + r \right) \, , &  c_4 =& \frac{4}{945} \left(6 + r \right) \, \left(30 - 23 r \right) \, . 
\end{aligned}
\end{split}
\ee
Neglecting the $k$-dependence in the regulator, the one-loop approximation of Eq.~\eqref{pdgl} is easily obtained
as
\be\label{pdgll}
\dot{\varphi}_k + 3\varphi_k - 2r\varphi'_k = \frac{2}{3 \pi^2}\left( 1 + \frac{r}{6}  \right)^\frac{3}{2}
\frac{\varphi'_k +  4(2-r) \varphi''_k}{3\varphi_k+4(1-r)\varphi'_k + 4\left(2-r\right)^2\varphi''_k}\, .
\ee
The l.h.s.~of these equations capture the classical scaling of the function $f_k(R)$, while the r.h.s.~originates from 
evaluating the operator trace of the FRGE and thus encodes the quantum corrections. The fixed
functions $\varphi_*(r)$ arising as regular, stationary solutions of \eqref{pdgl} and \eqref{pdgll} satisfy non-linear ordinary
differential equations (ODEs) of third and second order, respectively.

The first quantity of interest in the construction of fixed functionals is an estimate for the dimension of the space of regular solutions arising from these ODEs. Given Eqs.\ \eqref{pdgl} and \eqref{pdgll}, we expect that a solution is characterized by three (two) free parameters, respectively. When casting the ODE's into standard initial value form, one sees, however, that the r.h.s.~has poles at fixed values $r_{\rm sing}$ which correspond to roots of the denominator. The condition that the solution can be continued at $r_{\rm sing}$ then places non-local constraints on the initial values characterizing completely regular solutions.
Inspecting the one-loop and non-perturbative equations in the light of this structure reveals that even though the degrees of the equations are different, the  number of roots is adjusted accordingly.
Thus both equations have the same {\it index}, defined as the order of the equation minus the total number of fixed zeros in the denominator.
We take this as a further indication for the validity of the singularity counting argument \cite{Dietz:2012ic}.

We now focus on the asymptotic behavior of Eqs.~\eqref{pdgl} and \eqref{pdgll} in the IR-limit where $r \rightarrow \infty$.
This region can be accessed by applying singularity scaling techniques. In this course, we write the fixed point equations
in terms of the scaled variables $\rt = r \epsilon^{-\beta}$, $\beta > 0$ and $\vpt(\rt) = \epsilon^\alpha \varphi(r/\epsilon^\beta)$. 
Working in the limit $\epsilon \rightarrow 0$, the asymptotic region $r \rightarrow \infty$ is magnified and shifted to an area where 
$\rt$ and $\vpt(\rt)$ are both of order one.
Depending on the ratio of $\alpha$ and $\beta$ one identifies different asymptotic behaviors. For $\alpha < -5\beta/2$ ($\alpha < -3\beta/2$ in the one-loop case)
the IR-behavior is dominated by the classical l.h.s.~of the equations. In this case the quantum corrections decouple and one can easily solve the first order equations
obtaining the power-law behavior reported in the first line of Tab.~\ref{Tab1}. For $\alpha > -3\beta/2$ ($\alpha > -\beta/2$ in the one-loop case)
on the other hand the quantum effects on the r.h.s.~become dominant. The equation linearizes to a second order equation
which gives rise to the power-law solutions displayed in the last line of Tab.~\ref{Tab1}.
\begin{table}[t]
\tbl{\label{Tab1}Asymptotic behavior of the fixed functions $\varphi_*(r)$ of the one-loop equation \eqref{pdgll}
and the non-perturbative equation \eqref{pdgl} in the limit $r \rightarrow \infty$.
For classical and quantum dominance the equations linearize and admit power-law solutions.}
{\begin{tabular}{|c|c|c|}
\hline
          &    one-loop    &    non-perturbative \\ \hline
classical & $\varphi_*(r) \sim  \alpha_1\,  r^{3/2}$  & $\varphi_*(r) \sim  \alpha_1 \, r^{3/2}$  \\ \hline
balanced  & Eq.\ \eqref{firstorder1l} & Eq.\ \eqref{npsecond}  \\ \hline
\hspace{4mm} quantum \hspace{4mm}  & \hspace{4mm} $\varphi_*(r) \sim  \alpha_1 \, r^{5/4} + \alpha_3$ \hspace{4mm}   & \hspace{4mm} $\varphi_*(r) \sim \alpha_1 \, r^{3/2} + \alpha_2 \, r^{113/92} + \alpha_3$ \hspace{4mm} \\ \hline
\end{tabular}}
\end{table}
The interesting case is when the inequality bounding the classical case is actually saturated. In this case 
the classical and quantum contributions are balanced and the asymptotic regime is
described by the following fixed point equations
\be\label{balanced}
\begin{split}
{\rm 1l}: & \qquad 
 3\vpt_k - 2 \rt \vpt'_k = \frac{\rt^\frac{3}{2}}{9 \, \sqrt{6} \,  \pi^2} \, 
\frac{\vpt'_k -  4 \rt \vpt''_k}{3\vpt_k - 4 r \vpt'_k + 4 \rt^2 \vpt''_k}\, , \\
{\rm np}: & \qquad 
 3\vpt_k - 2 \rt \vpt'_k = \frac{\rt^\frac{5}{2}}{5670 \, \sqrt{6} \,  \pi^2} \, \frac{21 \vpt'_k + 50 \rt \vpt''_k + 184 \rt^2 \vpt'''_k   }{3\vpt_k - 4 r \vpt'_k + 4 \rt^2 \vpt''_k}\, , 
\end{split}
\ee
which arise from the one-loop and non-perturbative PDEs, respectively.
 
Since the Eqs.\ \eqref{balanced} encode the IR-behavior of our fixed functionals, it is worth to analyze them in more detail. Starting from the one-loop
approximation the substitution $\vpt(\rt) = (18 \, \sqrt{6} \pi^2)^{-1} \, \sqrt{z} \, (g(z)-1), \rt = z$ leads to 
\be\label{zoomed1l}
g_{zz} = - \frac{3}{8z^2} \, \frac{1 - g + 2 z g_z}{ g - z g_z } \, . 
\ee 
A priori, this equation looks like a non-linear ODE of second order. The structure of Eq.\ \eqref{zoomed1l} is
however special in the sense that it is of generalized homogeneous type, implying that the order of the equation is actually one. 
Applying the transformation $x = g^{-1}$, $y = z g^{-1} g'$, Eq.\ \eqref{zoomed1l} can be cast into the form
\be\label{firstorder1l}
x (y^2- y) \, y_x = - \tfrac{3}{8} (1-x) + y^3 - 2 y^2 + \tfrac{1}{4} y \, .
\ee
%
%
The non-perturbative equation can be recast in a similar fashion. Applying the transformation
$\vpt = 13 \, (1008 \, \sqrt{6} \, \pi^2)^{-1} \, z^{3/2} \, g(z), z = \rt$ we obtain
\be
g_{zzz} = - \frac{1}{184 z^2} \left( 1170\,  z \, g_z^2 + 878 \, z \, g_{zz} + 585 \, g_z \, (1 + z^2 \, g_{zz} ) \right) \, .  
\ee
Again, this equation is of generalized homogeneous type, so that the transformation $x = g$, $y = z g^{-1} g'$ reduces its order by one
\be\label{npsecond}
\begin{split}
- x^2 y y_{xx} = &    x^2 y_x^2 + (a_1 + 4 y + a_2 x y) x y_x
+ (1 + a_2 x) y^2   + a_2 x y + a_1 y + a_3 \, , 
\end{split}
\ee
where $a_1 = 163/92$, $a_2 = 585/184$, and $a_3 = 75/184$.
The balanced cases \eqref{firstorder1l} and \eqref{npsecond} have no power-law solutions.
Instead, a more general
asymptotic behavior has to be expected.
We believe that the newly found generalized homogeneous behavior is a rather generic feature
appearing in the IR-asymptotics of the fixed point equations.
Since generalized homogeneous equations are known to be effectively of one order less,
this implies that extra care is necessary when counting the free parameters of the solution in the balanced case.
%
\section{Conclusions}

The Asymptotic Safety program is currently on the verge of moving
from the exploration of finite-dimensional to infinite-dimensional ans{\"a}tze for the effective average action.
The $f(R)$-truncations of the form \eqref{truncation} thereby provide the simplest gravitational setting,
being the gravitational analogue to the local potential approximation in scalar field theory.
While there are still conceptual questions, as for example the continuation of \eqref{pdgl} to negative curvatures,
admissible boundary conditions, or the asymptotics of solutions at large values $r$,
awaiting their final answer we are positive that the NGFP found
on finite-dimensional truncation spaces has an admissible generalization to the realm of $f(R)$-gravity. 
\section*{Acknowledgements}
The research of F.S.\ and O.Z.\ is supported by the Deutsche Forschungsgemeinschaft (DFG)
within the Emmy-Noether program (Grant SA/1975 1-1).

\bibliographystyle{ws-procs975x65}

\end{document}